\documentclass[acmtog]{acmart}

\usepackage{booktabs} 
\usepackage{multirow}

\usepackage{url}

\acmJournal{TOG}
\acmVolume{38}
\acmNumber{4}
\acmArticle{91}
\acmYear{2019}
\acmMonth{7}

\acmDOI{0000001.0000001_2}

\definecolor{turquoise}{cmyk}{0.65,0,0.1,0.1}
\definecolor{purple}{rgb}{0.65,0,0.65}
\definecolor{dark_green}{rgb}{0, 0.5, 0}
\definecolor{orange}{rgb}{0.8, 0.6, 0.2}
\definecolor{red}{rgb}{0.8, 0.2, 0.2}
\definecolor{brown}{rgb}{0.5, 0.16, 0.16}

\newcommand{\rs}[1]{{\color{black}#1}}

\usepackage[ruled]{algorithm2e} 

\SetAlFnt{\small}
\SetAlCapFnt{\small}
\SetAlCapNameFnt{\small}
\SetAlCapHSkip{0pt}

\setcopyright{acmcopyright}

\begin{document}
\title{SAGNet: Structure-aware Generative Network for 3D-Shape Modeling}

\author{Zhijie Wu}
\affiliation{%
	\institution{Shenzhen University}
}
\author{Xiang Wang}
\affiliation{%
	\institution{Shenzhen University}
}
\author{Di Lin}
\affiliation{%
	\institution{Shenzhen University}
}
\author{Dani Lischinski}
\affiliation{%
	\institution{The Hebrew University of Jerusalem}
}
\author{Daniel Cohen-Or}
\affiliation{
	\institution{Shenzhen University and Tel Aviv University}
}
\author{Hui Huang}
\authornote{Corresponding author: Hui Huang (hhzhiyan@gmail.com)}
\affiliation{%
	\department{College of Computer Science \& Software Engineering}
	\institution{Shenzhen University}
}

\renewcommand\shortauthors{Z. Wu, X. Wang, D. Lin, D. Lischinski, D. Cohen-Or, and H. Huang}

\begin{abstract}

We present SAGNet, a structure-aware generative model for 3D shapes. Given a set of segmented objects of a certain class, the geometry of their parts and the pairwise relationships between them (the structure) are jointly learned and embedded in a latent space by an autoencoder. The encoder intertwines the geometry and structure features into a single latent code, while the decoder disentangles the features and reconstructs the geometry and structure of the 3D model. Our autoencoder consists of two branches, one for the structure and one for the geometry. The key idea is that during the analysis, the two branches exchange information between them, thereby learning the dependencies between structure and geometry and encoding two augmented features, which are then fused into a single latent code. This explicit intertwining of information enables separately controlling the geometry and the structure of the generated models. We evaluate the performance of our method and conduct an ablation study. We explicitly show that encoding of shapes accounts for both similarities in structure and geometry. A variety of quality results generated by SAGNet are presented. The data and code are at \url{https://github.com/zhijieW-94/SAGNet}. 

\end{abstract}


%
%

\setcopyright{acmcopyright}
\acmJournal{TOG}
\acmYear{2019}\acmVolume{38}\acmNumber{4}\acmArticle{91}\acmMonth{7} \acmDOI{10.1145/3306346.3322956}


\begin{teaserfigure}
	\centering
	\includegraphics[width=\textwidth]{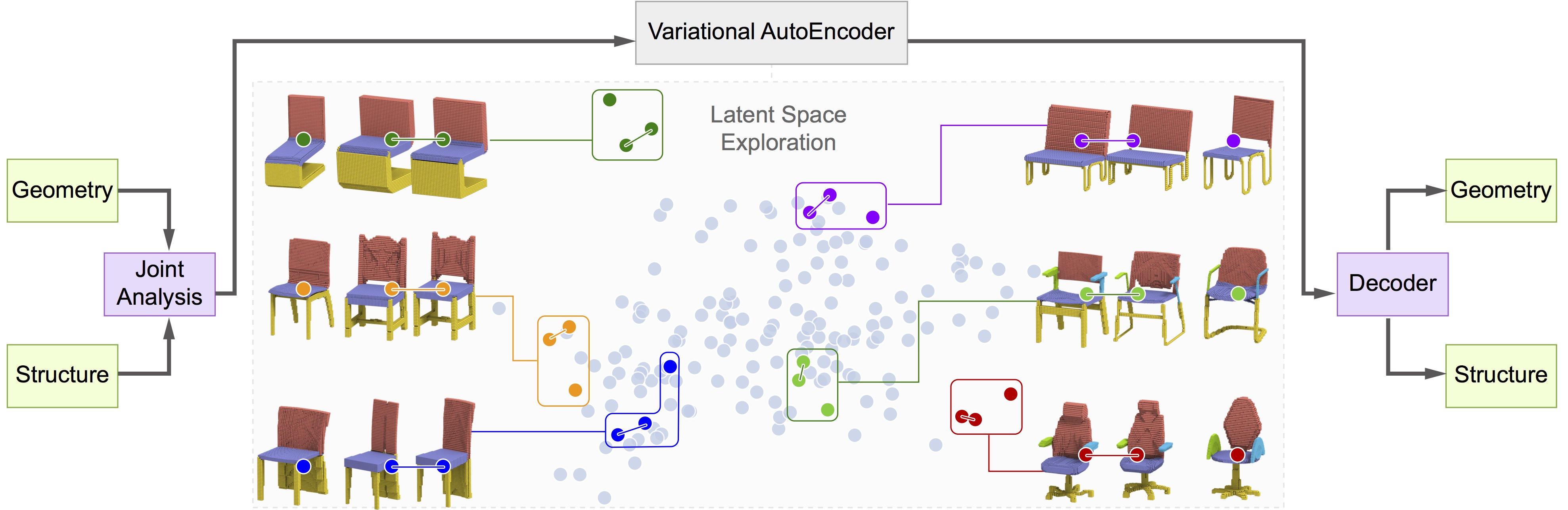}
	\caption{Our generative model jointly analyzes the structure and geometry of shapes, encoding them into a single latent code. The highlighted triplets above demonstrate that, in this joint latent space, pairs of nearby points represent models that are close to each other in both geometry and structure, while stepping away from the pair introduces differences in either geometry, or structure, or both. \rs{Note that all the shapes shown here are voxel-based, and the bounding boxes of their parts are hidden on purpose for a clearer visualization.}}
	\label{fig:teaser}
\end{teaserfigure}

\maketitle

\section{Introduction}

\begin{figure*}[t!]
	\centering
	\includegraphics[width=\linewidth]{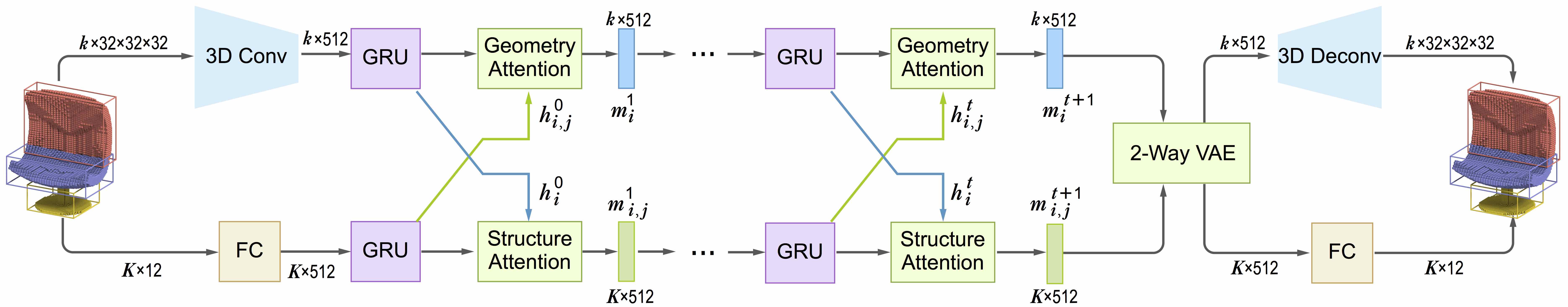}
	\caption{Overview of SAGNet. Given 3D shapes as training data, the network has traditional 3D convolutional and fully-connected layers to extract visual features for shape parts. The network is equipped with GRU-based encoder and attention component, which jointly analyzes the geometry and structural information of shapes. All the information are provided for the 2-way VAE, which offers the generative power to the network. Our network eventually decodes the geometry and structural information to generate 3D shapes.}
	\label{Fig:fig2}
\end{figure*}

Modeling of 3D shapes is a central problem in computer graphics. In recent years more attention has been given to structure-aware modeling techniques, where the relations among parts are carefully considered. Analyzing the structure provides a high-level understanding of the shape, and it goes beyond the low-level analysis of the local geometry~\cite{mitra2013}. The structure of a shape can be inferred from a single instance, but analyzing a family of shapes that share some similar structural characteristics can yield a much more powerful representation~\cite{Fish2014}. However, such an analysis is challenging, since structure and geometry are often inter-dependent, exhibiting complex relations and dependencies that are not easy to model directly.

Our work is motivated by recent advances in the competence of neural networks in analyzing data. We present a generative network that analyzes and encodes latent relationships between structure and geometry in a class of (man-made) shapes. \rs{Our framework leverages the parts of 3D shapes to learn discriminative representations. For a given set of shapes, structure is already implicitly represented in their geometric models, however, geometric and structural information is entangled together in a manner that does not provide a way to control each of them separately. Thus, instead of performing unsupervised training with geometric models of entire objects, we use a weakly supervised training strategy, where we provide the geometry as a collection of separate parts and the structure as pairs of their bounding boxes~\cite{mitra2013, li2017grass}. }

The high-level concept of our structure-aware generative network (SAGNet) is illustrated in Fig.~\ref{fig:teaser}. \rs{Nowadays, there are growing datasets of segmented shapes, such as~\cite{mo2018partnet, yi2016scalable}, and thus our network is trained using a set of shapes segmented into parts in order to leverage this readily available semantic information. All of the shapes are consistently oriented, and their parts are labeled.} The geometry of each part is provided as a voxel map and the structure is provided as a list of the bounding boxes, one for each part. Note that nothing is assumed about the relationships between the parts, thus the structure of the shape is not specified explicitly.
To jointly analyze the geometry and structure information of 3D shapes, we learn a latent space that encodes the relationship among all information into a code. In contrast to the conventional approaches that use a single code to implicitly represent all of the information, our latent code explicitly models their dependencies. Specifically, we exchange the geometry and structure information at two different stages. Before merging all the information, we model their dependencies to capture the lower-level information and build the joint structure-geometry latent space.
The learned latent space eventually benefits the reconstruction of both geometry and structure information. Moreover, the weak supervision allows to independently control geometry and structure when synthesizing new models.


As demonstrated in Fig.~\ref{fig:teaser}, nearby points in the joint latent space correspond to shapes that are similar to each other in both geometry and structure, while taking a step away introduces differences in geometry, structure, or both. Such latent space supports separate control of geometry and structure, and operations such as shape interpolation and completion.

The key novel component of our network is the exchange of data between the structure and geometry branches that takes place during the joint analysis stage.
Previous attempts in developing generative neural networks for 3D shapes, included adversarial networks based on voxels or point representations of the geometry~\cite{yan2016perspective,wu20153d,girdhar2016learning,choy20163d}, or structure-based approaches~\cite{li2017grass,zou20173d}. These methods do not leverage the power of a joint analysis of geometry and structure.
In this paper, we develop a novel architecture for joint analysis of geometry and structure information, which is necessary to address this challenging and difficult task. In particular, we focus on relationships between parts, which helps in learning the joint latent space.
We show that our generative network generates plausible structure-aware shapes that adhere to the characteristics of the learned class. We also demonstrate that our approach supports inferring the geometry from structure and vise-versa, enabling applications such as shape completion, and constrained modeling.




\section{Related Work}

In recent years there have been efforts to leverage the success of deep neural networks to develop generative models of 3D shapes. Wu et al.~\shortcite{wu20153d} develop a neural framework based on deep belief network to synthesize novel samples. Later, Wu et al.~\shortcite{wu2016learning} model the distribution of voxels of 3D objects using an adversarial approach. Their model can take a random noise as input and generate a voxel grid as output. The latent representation that they learn supports simple arithmetic and interpolation operations on the latent codes. Girdhar et al.~\shortcite{girdhar2016learning} embed voxel maps of shapes and their corresponding images in a shared latent space, making it possible to predicts a voxel map from a single 2D image. More advanced 3D generative voxel-based models are presented~\cite{yan2016perspective, gwak2017weakly, Tatarchenko2017OctreeGN}. \rs{Achlioptas et al.~\shortcite{achlioptas2017learning} introduce a generative adversarial network (GAN) in the latent space for point clouds}. Nash et al.~\shortcite{nash2017shape} present a generative model based on a variational autoencoder (VAE)~\cite{Kingma2014}. \rs{Yang et al.~\shortcite{Yang2017FoldingNetPC} apply a deformation-based approach to reconstruct point clouds. Sinha et al.~\shortcite{Sinha2017SurfNetG3} generate surfaces using deep neural networks, while Schor et al.~\shortcite{schor2018learning} propose to synthesize unseen shapes via part synthesis and composition.}

In our approach, we also employ a VAE to generate new shapes and use regular 3D voxel maps to represent their geometry. However, we represent each part using its own voxel map. Differently from the previous works above, we use recurrent neural networks (RNNs) to analyze the data. RNNs are widely used in generative models to analyze and generate novel sequences. For example, van den Oord et al.~\shortcite{oord2016pixel} regard natural images as sequences and generate images row by row, pixel by pixel. Rezende et al.~\shortcite{rezende2016unsupervised} develop a novel model based on DRAW~\cite{gregor2015draw}, which can reconstruct 3D structures from 2D images. Zou et al.~\shortcite{zou20173d} use an RNN decoder to generate shape primitives step by step from given depth images.

Our architecture consists of several RNNs and an RNN-based 2-way VAE that learns a generative representation for 3D shapes.
\rs{Unlike prior part-based generative models~\cite{Alhashim2014Topologyvarying3S, Huang2015AnalysisAS, G2L18}, we consider geometry and structure jointly, where the challenge is to learn a joint distribution of signals from different domains with neural networks~\cite{liu2017unsupervised,liu2016coupled}. } Other examples of methods that learn a joint distribution include Choy et al.~\shortcite{choy20163d}, who use 3D-R2N2 to build  a joint distribution to reconstruct 3D voxel maps from images. Also, Li et al.~\shortcite{li2015joint} get a joint embedding of images and 3D shapes via CNN purification. \rs{Tulsiani et al.~\shortcite{Tulsiani2017LearningSA} learn complex shapes with the mapping from voxel maps to corresponding primitives.}



Relationships among entities, or spatial layouts of objects, are known to be useful for understanding visual information. Some previous works explore the physical relationship~\cite{jia20133d,zheng2015scene}, while others~\cite{socher2011parsing,socher2012convolutional} use a recursive structure, and a recursive autoencoder, to capture the relationship by iteratively collapsing edges of a graph to yield a hierarchy.
Li et al.~\shortcite{li2017grass} adapt such recursive structures, and present a generative neural network model for the 3D structures of shapes, which can capture the structural information of different shapes within a class. Unlike us, they do not jointly consider the geometry domain and its corresponding structure domain, and they do not learn the dependencies between the geometries of different parts in an object.

\section{Structure-aware Generative Network}
\label{sec:overview}

\begin{figure}[t]
	\centering
	\includegraphics[width=\linewidth]{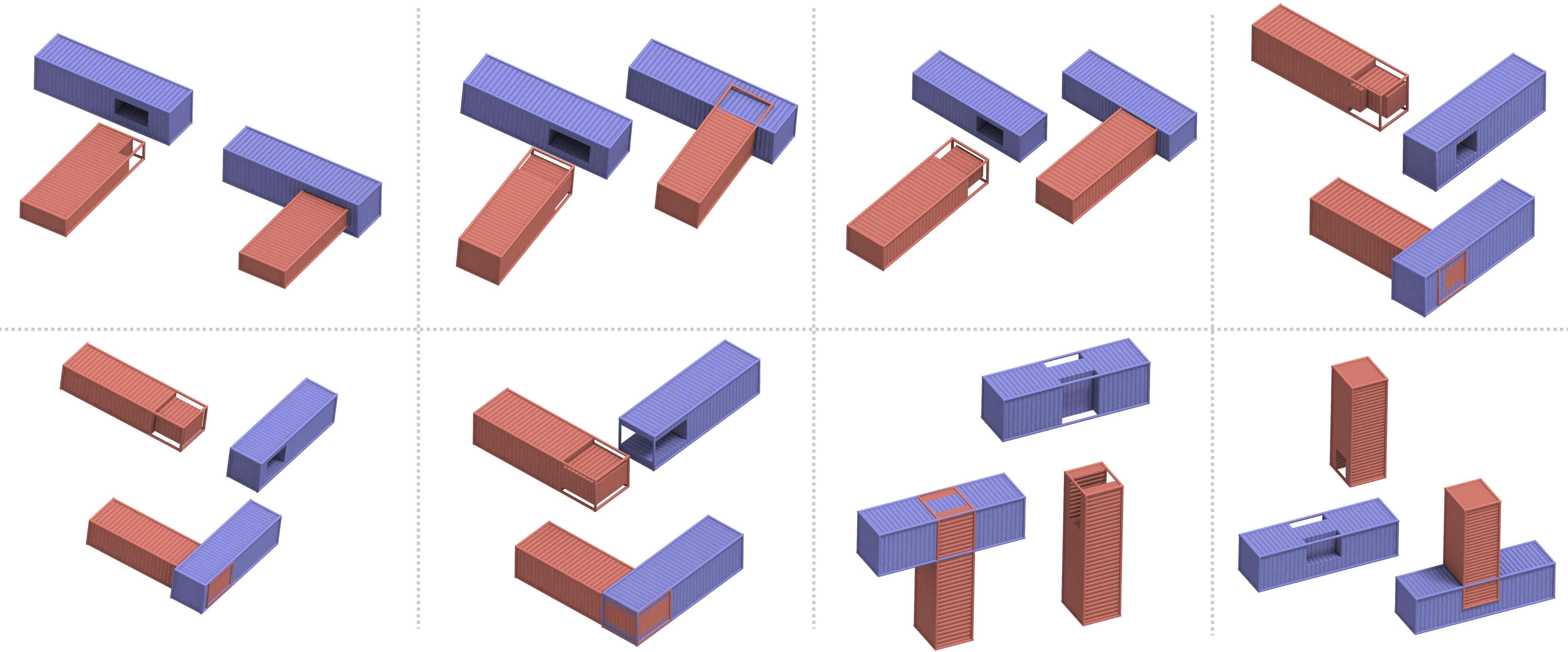}
	\caption{\rs{A class of two-part shapes shown above (often called as \emph{tenon-mortise joints}) represent eight different connectivities, where each pair indicates one connection mode and the red part fits tightly into the cavity of blue part.  Thus, it is challenging to infer the geometry of the blue part from its bounding box only, unless the relation to the red part is accurately learned. We take it as a training dataset; see results in Fig.~\ref{fig:toy}.} }
	\label{Fig:toy_example}
\end{figure}

The main idea of our method is to analyze and generate shapes by jointly considering their structure and geometry, learning them and their inter-relations.
\rs{
To demonstrate the importance of learning these inter-relations, consider a synthetic example of \emph{tenon-mortise joints} in Fig.~\ref{Fig:toy_example}. In this class of two-part shapes, the location and shape of the cavity inside the blue part is completely determined by the relative position and the shape of the red one. Thus, if the relation between the relative position of the parts and their geometry is not learned well by the network, it is unlikely that the network would succeed in generating the blue parts with a correctly sized and placed cavity. See also Fig.~\ref{fig:toy} for the comparison of our SAGNet with state-of-the-art methods on this dataset.
}

Following the conventional setting of GRASS~\cite{li2017grass} and the survey of Mitra et al.~\shortcite{mitra2013}, we define the structure based on a coarse segmentation instead of a fine annotation. To leverage the available semantic information and reduce the training complexity, our model is trained on a set of segmented and consistently oriented shapes.
Each shape is represented with $k$ parts, where each part consists of a bounding box that contains a voxel map representing the part geometry. We represent the shape structure as the set of all $K = k \times (k-1) / 2$ pairwise spatial relationships between the $k$ parts~\cite{mitra2013}. Thus, a shape is represented by two series (i) $k$ voxel maps, and (ii) $K$ pairs of axis-aligned bounding boxes, where each pair is represented by $2 \times 6$ coordinates. 
The first three coordinates specify the center of a box and the last three elements specify its length, width and height. The actual number of parts for a given class may be smaller than $k$, in which case the missing parts are represented using zeros. The ordering of the shape parts within the same class is consistent, as this makes the training easier. Note that the voxel map in each bounding box is processed independently. Thus, the overlap between the voxel grids of different parts does not affect the final output. The part order is uniform for each category of shapes, providing a consistent order for structure features. With this consistency, we build a correspondence between the structure features and different parts. It allows to associate (k-1) $6D$ candidate bounding boxes with one part. We average the (k-1) intermediate vectors for the final output of a bounding box.

Fig.~\ref{Fig:fig2} shows an overview of our architecture. Generally speaking, it is a two-branch autoencoder. The network takes two streams of input: one is a series of $k$ voxel maps (the upper branch in the figure), and the other is a series of $K$ pairwise spatial relationships, represented by pairs of bounding boxes (the lower branch). The geometry stream is analyzed by convolutional layers, and fed into an RNN component that analyzes the series of the resulting $k$ features. The structural stream is analyzed by fully-connected layers and fed into an RNN that analyzes the series of $K$ pairwise relation features.
Each of these RNN units is implemented using a Gated Recurrent Unit (GRU)~\cite{cho2014properties}.
Compared to other variants of recurrent neural networks (e.g., LSTM), GRU has fewer gates, but enables storing and filtering information with its internal memory. This alleviates the gradient vanishing problem during the network training. Furthermore, GRU has fewer learnable parameters, 
thereby achieving greater flexibility in a more principled training framework~\cite{chung2014empirical, cho2014properties}.

The outputs of these two GRUs are then fed into modules that exchange information between the geometry and structure streams. The exchange of information is weighted by the influence of the respective data. This is commonly called ``attention'': In the Geometry Attention module, the geometric features are given attention by their $k-1$ related structural features. Likewise, in the Structure Attention module, each structure feature is given attention by its two geometry features. Each attention module yields a deep feature, one representing the geometry and one for the structure. \rs{After few iterations, the pairwise relation representations will contain global context and have strong ability to model complex structural dependencies in shapes. More details are provided in Section~\ref{sec:details}}.


These two deep features are next fed into a 2-way VAE, which accepts two inputs (one from each branch), rather than one. The purpose of the 2-way VAE is to combine and and fuse the two features representing the geometry and the structure into a single vector, thereby embedding them in a joint latent space. On the output end, the 2-way VAE produces two features, associated with the two streams, which are fed into two corresponding branches with two decoders that generate the output streams.

\begin{figure}[t!]
	\centering
	\includegraphics[width=0.91\linewidth]{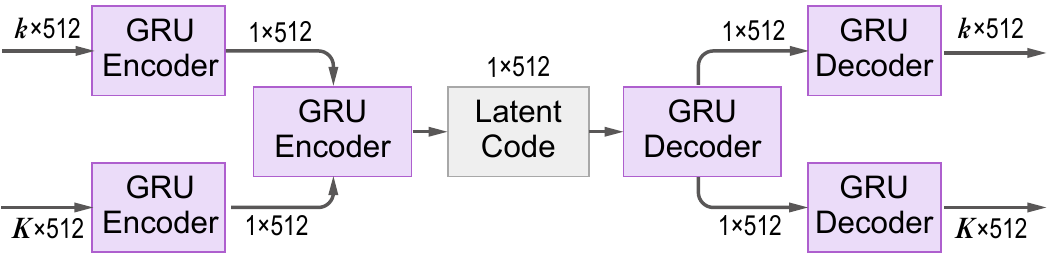}
	\caption{Architecture of 2-Way VAE. Our VAE has internal GRU encoders, which takes input as the geometry and structural information analyzed by the attention component. To generate a 3D shape, the encoders output a latent code that is processed by the internal GRU-decoders of the VAE.}
	\label{Fig:2VAE}
\end{figure}

The architecture of the 2-way VAE is depicted in Fig.~\ref{Fig:2VAE}. First, the two input feature streams are fed into corresponding GRU units whose goal is to collapse each feature stream into a single feature vector of size $1 \times 512$. The two resulting features are next fed into another GRU encoder, which fuses them into a single latent code. This latent code represents the coordinates of the shape in the joint embedding space. The fused features represent the shapes, encapsulating their geometries and structures.
In other words, the fused features are structure-aware since they encode the geometry and structure information, as well as the relation between them. The decoding end of the 2-way VAE exactly mirrors the encoding end, and consists of GRU decoders that split the features into two streams, which are then decoded into geometric and structural series.

Note that information from the geometry and structure streams is exchanged in two stages of the pipeline. First, during their analysis, using the Geometry and Structure attention modules, and then in the 2-Way VAE, where the two streams collapsed and fused.

\begin{figure*}[t]
	\centering
	\includegraphics[width=0.96\linewidth]{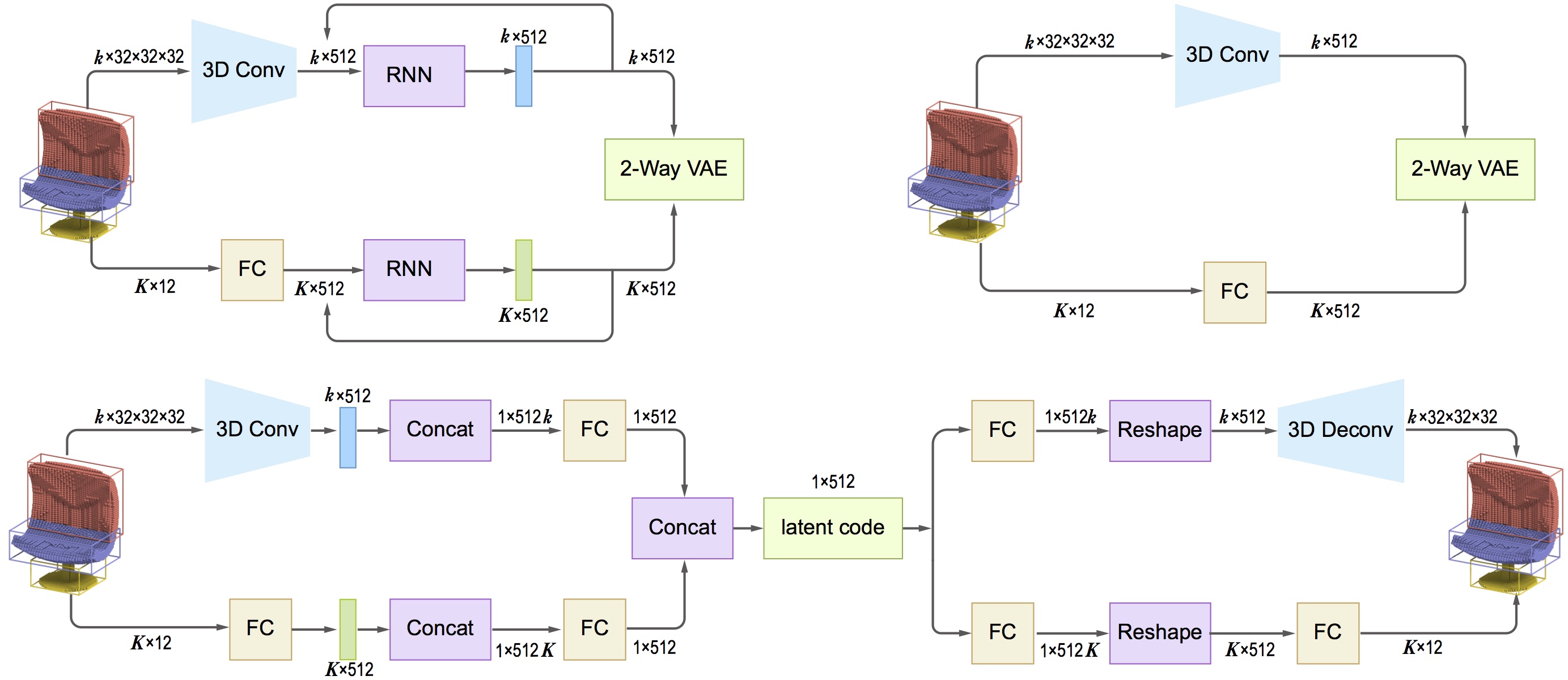}
	\caption{\rs{The architectures of three ablation study baseline models. The top left diagram denotes the \emph{No-attention} baseline model. The top right diagram corresponds to \emph{No-GRU} baseline model. The diagram that lies at bottom indicates the baseline model of \emph{Concatenation}.}}
	\label{fig:ablation_baseline}
\end{figure*}

\rs{To show the importance of the inclusion of GRUs and attention modules, we perform an ablation study on three simpler baselines, illustrated in Fig.~\ref{fig:ablation_baseline}, respectively. These three baseline networks use simpler schemes to analyze the geometry and structure information.
For a fair comparison, we only change the key component of one baseline model and keep other settings fixed.
The first called \emph{No-attention} baseline is defined by removing the attention component to disable the information exchange between the geometry and structure branches of SAGNet, as shown in the top left corner. The second baseline named \emph{No-GRU} is defined by removing the GRUs from both of these branches, as shown in the top right corner. Note that the name \emph{No-GRU} means that there is no GRUs in the auto-encoder but there are still GRUs in 2-way VAE. The removal of the GRUs, effectively disables the modeling of the information propagation between parts. The third variant is the \emph{Concatenation} baseline, built by replacing the six GRUs in the 2-way VAE with fully-connected layers and concatenation (abbr. concat) operation, as shown in the bottom row. This replacement is used to indicate that GRU is a better choice to fuse features as it is good at capturing dependencies between parts.}

\section{Implementation}
\label{sec:details}

In this section,  we describe the implementation of our approach and the training of our model in more detail.

\subsection{Two-Branch Autoencoder}


Fig.~\ref{Fig:fig2} shows the two-branch autoencoder, whose upper branch is intended for processing the geometry, while the lower branch processes the structure.
The geometry branch consists of five 3D convolutional layers on the encoder side, accepting a series of $k$ $(32 \times 32 \times 32)$ voxel maps as input. The 3D convolutional layers down-sample the voxel maps by a ratio of 16 and are followed by a fully-connected layer to compute $k$ $512$D features. In parallel, the structure branch has a fully-connected layer to process $K$ pairs of bounding boxes, producing $K$ $512$D features.

The features output by the encoder are fed into two different GRUs. These GRUs account for the relationships between parts in terms of geometry and structure, exchange information between them using the {\it Geometry and Structure Attention} components, and eventually output the $k$ $512$D features and $K$ $512$D features to the {\it 2-Way VAE}. Finally, the decoder echoes the encoder with five 3D deconvolutional layers that transform latent features back into voxel maps. It also has a fully-connected layer to regress latent features to $k$ bounding boxes for all parts.

\subsection{Geometry and Structure Attention Component}

The geometry and structure attention components are used to exchange the information between the upper and lower branches. The attention components are implemented with two fully-connected layers~\cite{xu2017scene}. We formulate the information exchanged between the the upper and lower branches as:
\begin{equation}
m^{t+1}_{i} = \sum_{j \neq i}f([h^{t}_{i}, h^{t}_{i,j}])h^{t}_{i,j},
\label{eq:msg_geo}
\end{equation}
\begin{equation}
m^{t+1}_{i,j}=f([h^{t}_{i,j}, h^{t}_{i}])h^{t}_{i}+f([h^{t}_{i,j}, h^{t}_{j}])h^{t}_{j},
\label{eq:msg_str}
\end{equation}
where $m^{t+1}_{i}$ and $m^{t+1}_{i,j}$ are the feedback messages for updating the hidden state of the upper- and lower-branch GRUs, respectively. $h^{t}_{i} \in \mathbb{R}^{512}$ is a geometry feature for the $i$-{th} part, which is produced by the upper-branch GRU. $h^{t}_{i,j}$ is the structure feature between the $i$-{th} and $j$-{th} parts, which is produced by the lower-branch GRU (see Fig.~\ref{Fig:fig2}). $t$ indicates the $t$-{th} iteration of GRUs. $f$ represents fully-connected sigmoid-activated layers. To simplify notations, we denote all full-connected layers as $f$.

The feedback message $m^{t+1}_{i}$ is computed as follows. For the $i$-{th} object part, we concatenate $h^{t}_{i}$ with a structure feature $h^{t}_{i,j}$, feeding them to a fully-connected layer with sigmoid activation function. The outcome feature of the fully-connected layer weights the structure feature $h^{t}_{i,j}$, attending to the relevant components of $h^{t}_{i,j}$ that can be employed by the upper-branch GRU to update the geometric feature $h^{t+1}_{i}$. The message $m^{t+1}_{i,j}$ is computed by two terms. Using the structural feature $h^{t}_{i,j}$, the first term models the attention of the geometry feature $h^{t}_{i}$, while the second term attends to the geometry feature $h^{t}_{j}$. Eq.~\eqref{eq:msg_str} summarizes the geometry information of the $i$-{th} and $j$-{th} parts for updating their structural feature $h^{t+1}_{i,j}$.

Using the messages defined in Eq.~\eqref{eq:msg_geo} and~\eqref{eq:msg_str}, we exchange the features representing the geometry and structure information between the two branches of SAGNet, as shown in Fig.~\ref{Fig:fig2}. It is crucial to generate accurate 3D object models, which rely on the joint information of geometry and structure. We omit the superscript $t$ in the last iteration of GRUs, denoting the geometry feature as $h_{i}$ for the $i$-{th} part and the structural feature $h_{i,j}$ for the $i$-{th} and $j$-{th} parts. The resulting geometry and structural features are fed into the 2-way VAE, which is elaborated below.

\begin{figure*}[t]
	\centering
	\includegraphics[width=\linewidth]{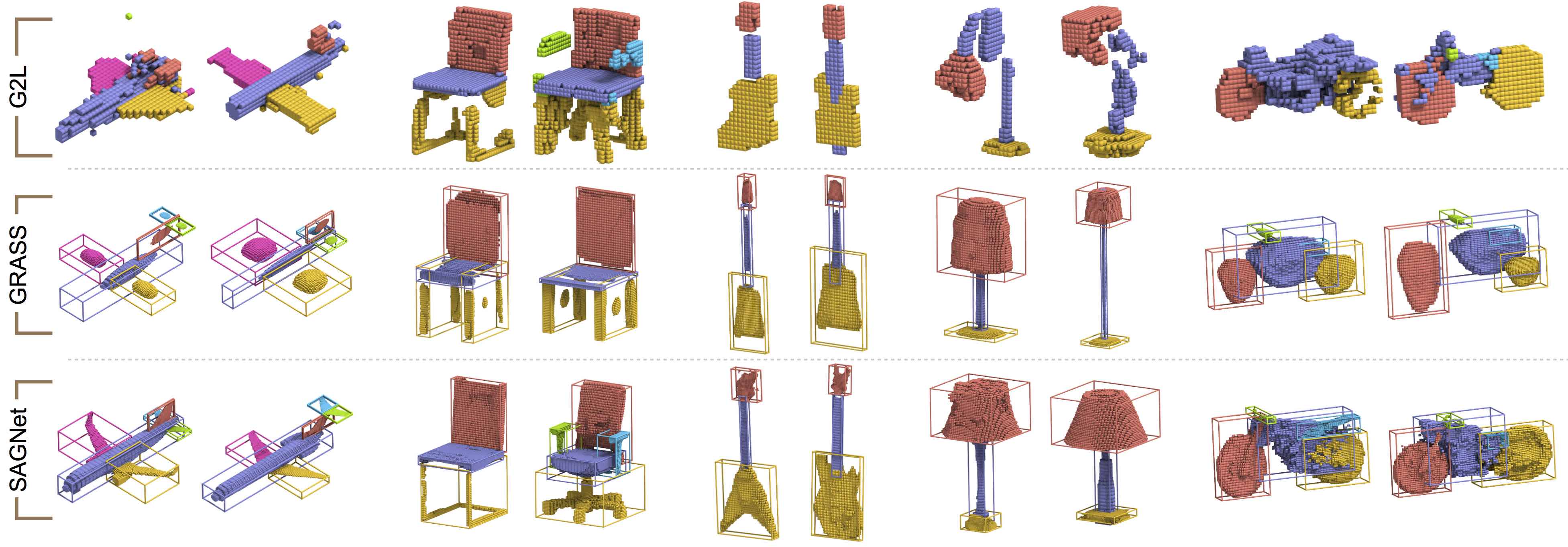}
	\caption{\rs{The generation results of G2L~\cite{G2L18} (top row), GRASS~\cite{li2017grass} (middle row) and SAGNet (bottom row). Compared to SAGNet, G2L often generates coarser and less-structured shapes while the voxel grids of GRASS may lose geometric details and thus are less visually appealing.}}
	\label{fig:G2L_Grass_gen}
\end{figure*}

\subsection{2-Way VAE}


The 2-way VAE also has an encoder-decoder architecture, which focuses on learning the dependencies between the geometry and structural features of a shape. As shown in Fig.~\ref{Fig:2VAE}, the 2-way VAE has an internal encoder, which consists of three GRUs. One GRU takes a sequence of geometry features $H_{g}=\{h_i | i=1,...,k\}$ (i.e., the $k$ $512$D features) as input. A second GRU processes the sequence of structural features $H_{s}=\{h_{i,j} | i=1,...,k, j>=i\}$ (i.e., the $K$ $512$D features). Each of these two GRUs encodes its input feature sequence into a single $512$D feature as:
\begin{equation}
h_{g} = G^{e}_{g}(H_{g}),~~~h_{s} = G^{e}_{s}(H_{s}),
\label{eq:hg}
\end{equation}
where $G^{e}_{g}$ and $G^{e}_{s}$ represent GRUs. Thus, $h_{g}, h_{s} \in \mathbb{R}^{512}$ encodes the global geometry and structural information of all parts, respectively.

Since different shapes in the analyzed family may consist of different subsets of the $k$ parts, in order to reconstruct such shapes it is necessary to provide a part mask
$c \in \mathbb{R}^{k}$. Each element of $c$ is a binary variable that indicates the presence (1) or absence (0) of a part. Using the part mask $c$ together with the geometry and structural features $h_{g}$ and $h_{s}$, we produce a new joint feature $h_{v} \in \mathbb{R}^{512}$ as:
\begin{equation}
h_{v} = G^{e}_{v}(f([h_{g}, c]), f([h_{s}, c])),
\label{eq:part_mask}
\end{equation}
where $G^{e}_{v}$ is a third GRU in the encoder of the 2-way VAE. The new feature $h_{v}$ passes through an extra fully-connected layer to yield two 512D vectors, which are the mean and standard deviation of a Gaussian distribution. We then generate a random variable $n \in \mathbb{R}^{512}$ to produce a latent vector $z \in \mathbb{R}^{512}$ as:
\begin{equation}
z = \mu + \sigma n,
\label{eq:random_sample}
\end{equation}
where $\mu$ and $\sigma$ represent the mean and standard deviation.

The 2-way VAE also has an internal decoder that processes the latent vector $z$. Again, the decoder has three GRUs. Following the decoding procedure of~\cite{roberts2017hierarchical,bowman2015generating}, $z$ is input to a GRU that outputs two 512D features. One 512D feature is fed to a decoder GRU to generate $k$ $512$D geometry features, and another 512D feature is used by another decoder GRU to produce $K$ $512$D structural features. Finally, the geometry and structural features are further processed by the decoder of our two-branch autoencoder to produce voxel maps and corresponding bounding boxes for all parts, as already explained earlier. 



\begin{figure*}[t!]
	\centering
	\includegraphics[width=\linewidth]{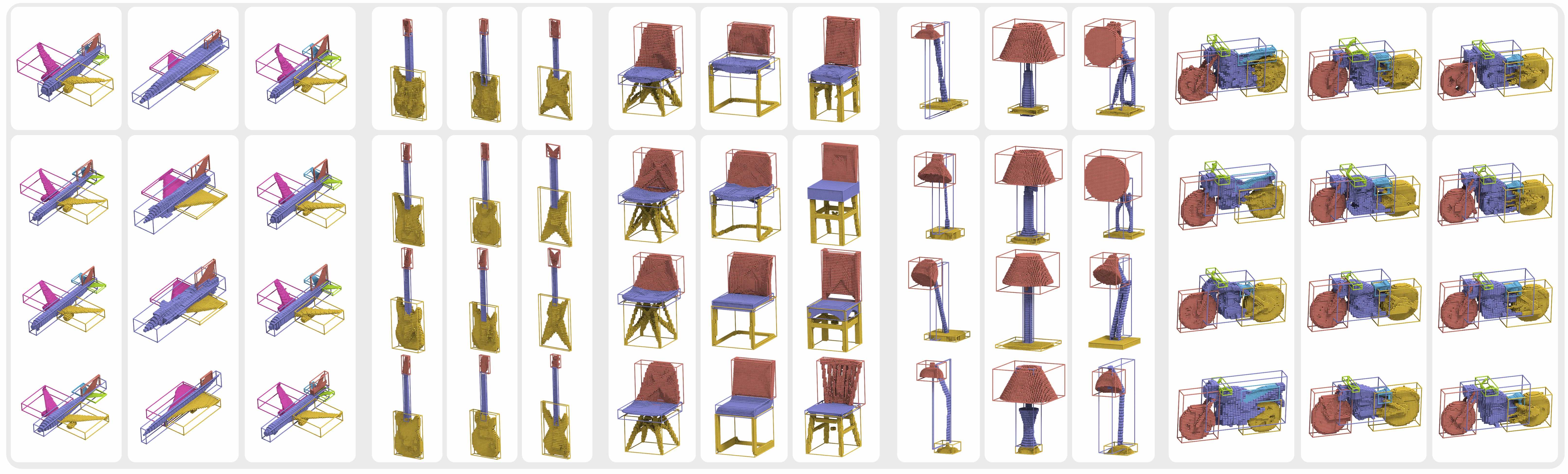}
	\caption{For each generated sample in the top row, we retrieve the 3-nearest neighbors in the training data. It may be seen that the generated shapes are original.}
	\label{Fig:knn_retrieval}
\end{figure*}

\subsection{Network Training}

The training of SAGNet includes two phases. In the first phase, we use a reconstruction loss to guide the training of the whole two-branch autoencoder. The first phase warms up the network training, avoiding the posterior collapse problem of the  VAE~\cite{bowman2015generating,shen2018improving}. In the second phase, we keep the reconstruction loss for the two-branch autoencoder, while adding a KL loss and feature regularization for the training of our 2-way VAE. We detail the two phases below.

In the first phase, we define the training objective function as:
\begin{equation}
L_f = -E_{q_{\phi}(z|v, b, c)}[log(p_{\varphi}(v, b|z,c))],
\label{eq:first}
\end{equation}
where $v$ and $b$ denote the voxel maps and bounding boxes. $c$ is the part mask that indicates the presence/absence of parts. $z$ is the latent feature produced in the 2-way VAE. The distribution $q_{\phi}(z|v, b, c)$ is output by the encoder part of 2-way VAE, and the distribution $p_{\varphi}(v, b|z,c)$ is output by our two-branch autoencoder. The objective function $L_f$ penalizes the reconstruction loss of the voxel maps and bounding boxes using the latent vector $z$.

In the second phase, we define the training objective as:
\begin{equation}
L_s = L_f + \lambda L_{KL} + \eta R,
\label{eq:second}
\end{equation}
where
\begin{equation}
L_{KL} = KL(q_\phi(z|x, y, c)||p_\phi(z|c)),
\label{eq:KL}
\end{equation}
\begin{equation}
R = \sum^{k}_{i=1}||h'_i - h_i||^{2}_{2} + \sum^{k}_{i=1}\sum^{k}_{j=i+1}||h'_{i,j} - h_{i,j}||^{2}_{2},
\label{eq:regularization}
\end{equation}
where $p_\phi(z|c)$ is a standard Gaussian distribution as prior, $h'_i$ denotes the $i$-{th} part's geometry feature, and $h'_{i,j}$ denotes the structural feature of the $i$-{th} and $j$-{th} parts. Both $h'_i$ and  $h'_{i,j}$ are produced by the 2-way VAE.
During the second phase, we gradually increase the factors $\lambda$ and $\eta$ in a linear manner from 0 to 0.8 within 60000 iterations, which provides SAGNet its generation power.

We use the TensorFlow platform~\cite{abadi2016tensorflow} to construct SAGNet. All the parameters are randomly initialized and optimized by the standard SGD solver. The network is trained with a learning rate of 0.001 for 70000 mini-batches, using a mini-batch size of 10.

\section{Results and Evaluation}

\setlength{\tabcolsep}{7pt}
\renewcommand{\arraystretch}{1.2}
\begin{table}
	\caption{\rs{Classes and numbers of objects and parts in our shape dataset. The class of tenon-mortise joints is synthesized to evaluate the necessity of joint analysis of geometry and structure. 10000 joints are generated for training to enhance the data diversity and thus prevent from overfitting.}}
	\small
	\begin{center}
		\begin{tabular}{c||cccccc}
			\hline
			class      &joint & airplane  &  chair  & guitar &  lamp  & motor   \\
			\hline\hline
			\# objects &  10000 & 1605      &  1384   & 779    &  1000  & 202          \\
			\# parts   &  2 & 6         &  7      & 3      &  4     & 5            \\
			\hline
		\end{tabular}
	\end{center}
	\vspace{-0.15in}
	\label{tab:dataset_info}
\end{table}


\begin{figure}[t!]
	\centering
	\includegraphics[width=\linewidth]{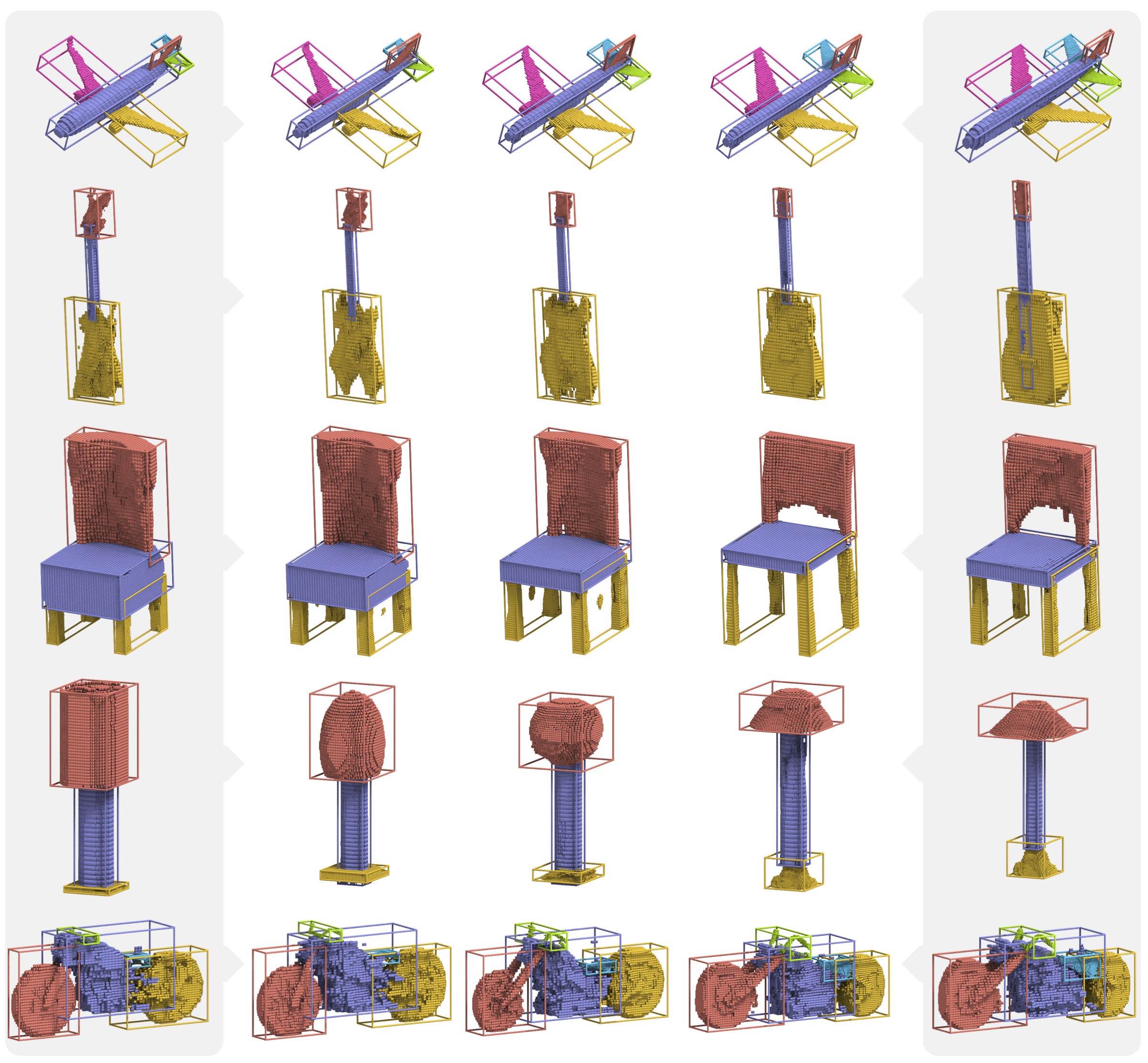}
	\caption{The left- and right-most shapes are randomly selected and paired from the training data. Their latent codes are then linearly interpolated to create three intermediate latent codes that are used to generate the shapes shown in the three middle columns.}
	\label{Fig:linear_interpolation}
\end{figure}

\begin{table*}[]
	\caption{\rs{Comparisons between SAGNet and baselines and existing models on quantitative scores over five classes of man-made shapes. In this table,  \emph{Completion} indicates the shape completion level. G2S and S2G denote geometry-to-structure mapping and structure-to-geometry mapping, respectively. The MMD and COV scores are computed following~\cite{achlioptas2017learning}. The smaller MMD score indicates the generated results have better fidelity. The higher COV score shows that one model can generate more diverse results. Our SAGNet achieves the best results in most cases. It demonstrates that the joint analysis of geometry and structure improves the generative quality in terms of fidelity and diversity.}}
	\begin{tabular}{clclccccccc}
		\hline
		\multicolumn{2}{c}{Dataset}                    & \multicolumn{2}{c}{Models}        & MMD-CD          & COV-CD         & MMD-EMD         & COV-EMD        & Completion      & G2S             & S2G             \\ \hline\hline
		\multicolumn{2}{c}{\multirow{6}{*}{Chair}}     & \multicolumn{2}{c}{SAGNet}        & \textbf{0.0024} & 0.751          & \textbf{0.0608} & 0.743          & \textbf{0.0600} & 0.0581 & \textbf{0.0258}         \\ \cline{3-11}
		\multicolumn{2}{c}{}                           & \multicolumn{2}{c}{No-attention}  & 0.0025          & 0.750          & 0.0612          & 0.760          & 0.0618          & \textbf{0.0579}          & 0.0271 \\ \cline{3-11}
		\multicolumn{2}{c}{}                           & \multicolumn{2}{c}{No-GRU}        & 0.0025          & 0.705          & 0.0620          & 0.746          & 0.0625          & 0.061         & 0.0264          \\ \cline{3-11}
		\multicolumn{2}{c}{}                           & \multicolumn{2}{c}{Concatenation} & 0.0029          & 0.665          & 0.0640          & 0.690          & 0.0645              & 0.0652             & 0.0356              \\ \cline{3-11}
		\multicolumn{2}{c}{}                           & \multicolumn{2}{c}{G2L}           & 0.0034          & \textbf{0.837} & 0.0682          & \textbf{0.834} & -               & -               & -               \\ \cline{3-11}
		\multicolumn{2}{c}{}                           & \multicolumn{2}{c}{GRASS}         & 0.0030          & 0.460          & 0.0744          & 0.445          & -               & -               & -               \\ \hline\hline
		\multicolumn{2}{c}{\multirow{6}{*}{Airplane}}  & \multicolumn{2}{c}{SAGNet}        & \textbf{0.0003} & \textbf{0.852} & \textbf{0.0247} & \textbf{0.889} & 0.0240          & \textbf{0.0265} & \textbf{0.0051} \\ \cline{3-11}
		\multicolumn{2}{c}{}                           & \multicolumn{2}{c}{No-attention}  & 0.0004          & 0.823          & 0.0252          & 0.854          & \textbf{0.0239} & 0.0269 & \textbf{0.0051}          \\ \cline{3-11}
		\multicolumn{2}{c}{}                           & \multicolumn{2}{c}{No-GRU}        & 0.0005          & 0.782          & 0.0259          & 0.796          & 0.0269          & 0.0282         & 0.0053         \\ \cline{3-11}
		\multicolumn{2}{c}{}                           & \multicolumn{2}{c}{Concatenation} & 0.0005          & 0.826          & 0.0255          & 0.845          & 0.028              & 0.032             & 0.012              \\ \cline{3-11}
		\multicolumn{2}{c}{}                           & \multicolumn{2}{c}{G2L}           & 0.0014          & 0.713          & 0.0310          & 0.797          & -               & -               & -               \\ \cline{3-11}
		\multicolumn{2}{c}{}                           & \multicolumn{2}{c}{GRASS}         & 0.0015          & 0.321          & 0.0472          & 0.328          & -               & -               & -               \\ \hline\hline
		\multicolumn{2}{c}{\multirow{6}{*}{Lamp}}      & \multicolumn{2}{c}{SAGNet}        & \textbf{0.0020} & \textbf{0.883} & 0.0563          & 0.874          & \textbf{0.0649} & \textbf{0.0824} & \textbf{0.0140} \\ \cline{3-11}
		\multicolumn{2}{c}{}                           & \multicolumn{2}{c}{No-attention}  & 0.0022          & 0.876          & 0.0580          & \textbf{0.886} & 0.0674          & 0.0853         & 0.0141          \\ \cline{3-11}
		\multicolumn{2}{c}{}                           & \multicolumn{2}{c}{No-GRU}        & 0.0022          & 0.869          & 0.0584          & 0.860          & 0.0650          & 0.0832          & 0.0161          \\ \cline{3-11}
		\multicolumn{2}{c}{}                           & \multicolumn{2}{c}{Concatenation} & 0.0025          & 0.827          & 0.0583          & 0.828          & 0.0795              &  0.0924             & 0.0331             \\ \cline{3-11}
		\multicolumn{2}{c}{}                           & \multicolumn{2}{c}{G2L}           & 0.0024          & 0.851          & \textbf{0.0510} & 0.873          & -               & -               & -               \\ \cline{3-11}
		\multicolumn{2}{c}{}                           & \multicolumn{2}{c}{GRASS}         & 0.0026          & 0.747          & 0.0591          & 0.764          & -               & -               & -               \\ \hline\hline
		\multicolumn{2}{c}{\multirow{6}{*}{Guitar}}    & \multicolumn{2}{c}{SAGNet}        & \textbf{0.0004} & \textbf{0.905} & \textbf{0.0300} & 0.906          & \textbf{0.0066} & 0.0238 & \textbf{0.0038}         \\ \cline{3-11}
		\multicolumn{2}{c}{}                           & \multicolumn{2}{c}{No-attention}  & \textbf{0.0004} & 0.902          & 0.0303          & \textbf{0.919} & 0.0067          & 0.0242           & 0.0039         \\ \cline{3-11}
		\multicolumn{2}{c}{}                           & \multicolumn{2}{c}{No-GRU}        & \textbf{0.0004} & 0.901          & 0.0303          & 0.911          & 0.0067          & \textbf{0.0237}         & 0.0039 \\ \cline{3-11}
		\multicolumn{2}{c}{}                           & \multicolumn{2}{c}{Concatenation} & 0.0005          & 0.877          & 0.0303          & 0.892          & 0.0277              & 0.0324              & 0.013              \\ \cline{3-11}
		\multicolumn{2}{c}{}                           & \multicolumn{2}{c}{G2L}           & 0.0100          & 0.569          & 0.0721          & 0.614          & -               & -               & -               \\ \cline{3-11}
		\multicolumn{2}{c}{}                           & \multicolumn{2}{c}{GRASS}         & 0.0007          & 0.551          & 0.0357          & 0.6573         & -               & -               & -               \\ \hline\hline
		\multicolumn{2}{c}{\multirow{6}{*}{Motor}} & \multicolumn{2}{c}{SAGNet}        & \textbf{0.0006} & \textbf{1.0}   & 0.0336          & \textbf{1.0}   & \textbf{0.0056} & \textbf{0.0350} & \textbf{0.0090} \\ \cline{3-11}
		\multicolumn{2}{c}{}                           & \multicolumn{2}{c}{No-attention}  & 0.0007          & \textbf{1.0}   & 0.0339          & 0.995          & 0.0058          & 0.0365          & 0.0092         \\ \cline{3-11}
		\multicolumn{2}{c}{}                           & \multicolumn{2}{c}{No-GRU}        & 0.0007          & \textbf{1.0}   & 0.0336          & \textbf{1.0}   & 0.0059          & 0.0365          & 0.0094          \\ \cline{3-11}
		\multicolumn{2}{c}{}                           & \multicolumn{2}{c}{Concatenation} & 0.0008          & 0.995          & 0.0346          & 0.995          & 0.0207              & 0.038             & 0.022              \\ \cline{3-11}
		\multicolumn{2}{c}{}                           & \multicolumn{2}{c}{G2L}           & 0.0007          & 0.995          & \textbf{0.0237} & 0.995          & -               & -               & -               \\ \cline{3-11}
		\multicolumn{2}{c}{}                           & \multicolumn{2}{c}{GRASS}         & 0.0027          & 0.505          & 0.0573          & 0.599          & -               & -               & -               \\ \hline
	\end{tabular}
	\label{tab:total_score}
\end{table*}

\rs{In addition to the synthetic example of tenon-mortise joints (Fig.~\ref{Fig:toy_example}), we also evaluate SAGNet using the data collected by Yi et al.~\shortcite{yi2016scalable} and Kae et al.~\shortcite{kae2013augmenting}, which contain part-labeled 3D shapes. There are five classes of objects, including airplane, chair, guitar, lamp and motor; see class and object details in Table~\ref{tab:dataset_info}.
Note that, since some of the original training shapes are noisy and segmented or labeled incorrectly, we re-segment the problematic parts manually, and then use these consolidated shape parts to generate voxel maps and bounding boxes as our input. 


We compare our SAGNet to state-of-the-art methods, e.g., the most recent G2L~\cite{G2L18} and GRASS~\cite{li2017grass}. We employ G2L code package provided by the authors to generate their results. 
In GRASS the structure and geometry are processed at different stages and only the training codes to synthesize bounding boxes have been released. To closely follow the description of Li et al.~\shortcite{li2017grass} and perform a fair comparison, we adapt the implementation to our training data by recursively merging the parts into one root vector, using the same classic strategy proposed by Socher et al.~\shortcite{socher2011parsing}. 
We then develop the geometry synthesis module by setting up an auto-encoder architecture and transformation module as applied in GRASS.
Both methods are retrained using their original hyperparameters to generate the results reported here.}


\subsection{Shape Generation}
\label{subsec:generation}

We train a SAGNet model for each class of objects. Using the trained network, we can generate object shapes that are represented by voxel maps and bounding boxes for each of their parts. To generate an object shape, we first sample a 512D latent code from a standard Gaussian distribution, using Eq.~\eqref{eq:random_sample} that generates the latent code with the internal encoder of 2-way VAE. Next, we input the sampled latent code to the decoder of 2-way VAE, computing the voxel maps and bounding boxes of all parts. \rs{Please refer to our supplementary material for a gallery of generated shapes by SAGNet and its ablation variants. For each part of a generated shape, there is an associated voxel map and a bounding box.
	
A set of shapes generated by G2L, GRASS and SAGNet are shown in Fig.~\ref{fig:G2L_Grass_gen} row-by-row for a qualitative comparison. As may be seen, compared to SAGNet, the generated results of G2L are coarser and less structured. As for GRASS, since it processes geometry and structure in two separate stages, although the generated shape structures are visually satisfactory, their corresponding voxel grids often have artifacts with geometric detail loss. 
Notably, our experiments show that even when GRASS is fed with part label information following the classic strategy~\cite{socher2011parsing}, it still produces less satisfactory results compared to SAGNet. The reason is that, given the part labels, SAGNet is able to model the part relationships better, successfully allowing the semantic dependency of labels to better enhance each part's geometry and structure representations.}

The generative network creates novel parts for the shapes, which are different from the given training data. To show the generative power of SAGNet, we compare the generated shapes with the training data. For a generated shape, we retrieve its 3-nearest neighbors in the training data, where the distance between two shapes is the sum of Chamfer distances between corresponding parts~\cite{achlioptas2017learning}. As may be seen in Fig.~\ref{Fig:knn_retrieval}, the generated shapes are original, and exhibit various differences from their nearest neighbors.

The distances between two shapes are defined based on their bounding boxes and voxel maps. Given the bounding boxes, we compute and sum the Euclidean distances between the corresponding parts of two shapes. In the voxel maps of a shape, we employ the 3D coordinates of the associated bounding boxes to compute the point cloud for representing each occupied voxel. For each voxel of a shape, we compute the Euclidean distance to the nearest voxel of another shape. All the distances between voxels are summed to form a Chamfer distance. The Euclidean distance of bounding boxes and Chamfer distance of point clouds are summed again as the overall distance between two shapes.

We conjecture that SAGNet combines the existing patterns of the training data, creating new patterns for the generated shapes. To verify this, we select two different shapes in the same class from the training set. Their voxel maps and bounding boxes of parts are fed into SAGNet, which outputs two latent vectors corresponding to the given input shapes. Then we perform linear interpolation using the two latent vectors. By controlling the interpolation rate, we  compute the latent code of novel shapes, which can be regarded as the combination of the training data. The interpolated latent code is input to the decoder of 2-way VAE for generating voxel maps and bounding boxes of the interpolated shape. \rs{We show the randomly paired training shapes in left- and right-most columns of Fig.~\ref{Fig:linear_interpolation}, and the generated shapes in middle columns based on the computed latent vectors. A smooth transition from one training shape to another can be built with our SAGNet.}

\begin{figure*}[t!]
	\centering
	\includegraphics[width=0.95\linewidth]{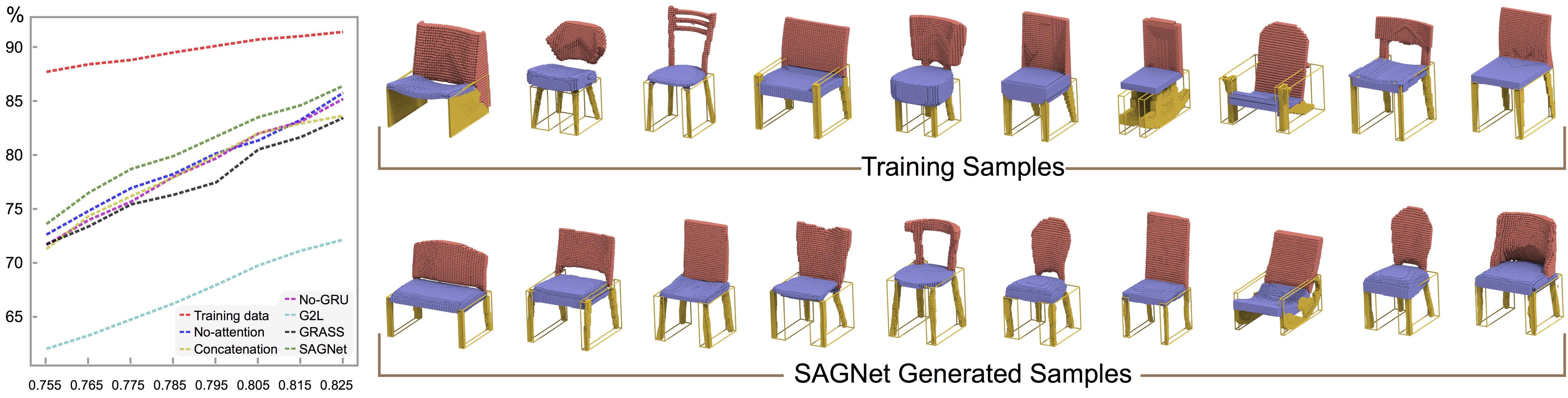}
	\caption{\rs{We measure the symmetry scores for legs of chairs generated by SAGNet, or baselines, or existing models. Along the horizontal axis, we set different thresholds for the scores. Along the vertical axis, we provide the percentage of shapes, which have smaller scores than the given thresholds. }}
	\label{Fig:symmetry_score}
\end{figure*}

\begin{figure*}[t!]
	\centering
	\includegraphics[width=0.95\linewidth]{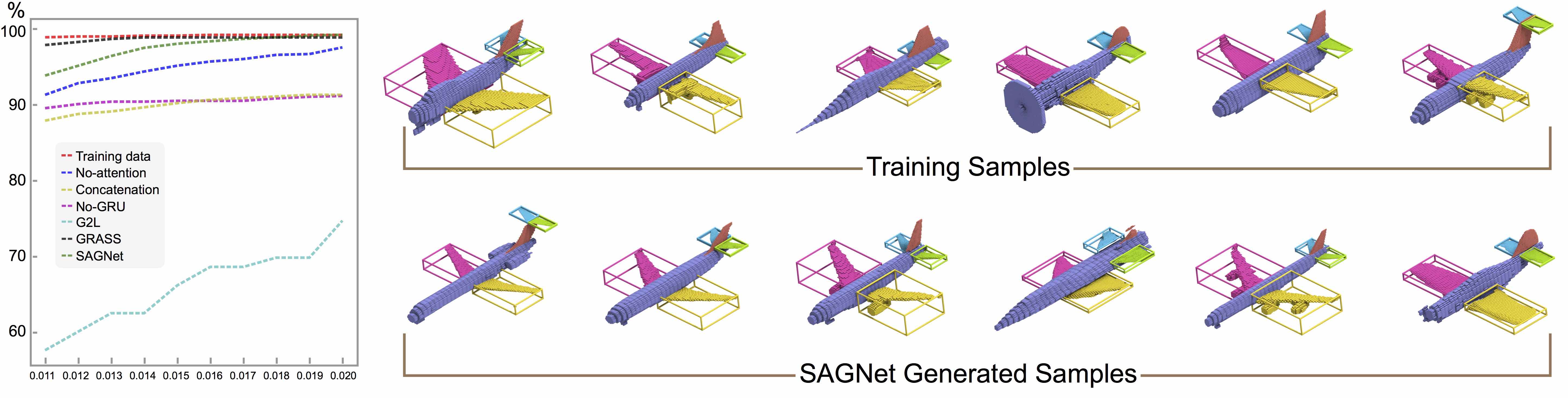}
	\caption{\rs{We measure the centroid-to-plane distances for each airplane generated by SAGNet, or baselines, or existing models. The centroid-to-plane distances are computed using the fore- and back-wings of airplanes. Along the horizontal axis, we provide different thresholds for the distances. Along the vertical axis, we provide the percentage of shapes, which have smaller distances than the given thresholds.}}
	\label{Fig:coplanarity_score}
\end{figure*}

We have also shown that vectors sampled from the latent space correspond to object shapes. Here, we further investigate the properties of the latent space. For better visualization, we retrain SAGNet by modifying the dimension of the latent space to two~\cite{nash2017shape}. This allows us to show and compare the latent vectors in a 2D Euclidean space. We find that similar shapes appear to be close in the 2D latent space, forming apparent clusters, as shown in Fig.~\ref{fig:teaser}. This demonstrates that SAGNet builds a proper relationship between the shapes and latent vectors, which is important for the shape generation task.


\rs{To evaluate the generative quality of all the models, we apply the MMD (Minimum Matching Distance) and COV (Coverage) metrics~\cite{achlioptas2017learning} on the generated shapes. These two metrics are used to evaluate the fidelity and diversity of a generative model, respectively. Note that, since one generated sample can be affected by parts from different training shapes, we evaluate MMD and COV with parts. To compute the COV score, we list all the generated parts, and match to the closest parts in the training data. Then we pick shapes which these matched parts belong to and calculate the fraction of the picked shapes to the shapes in training data as the COV score. A higher coverage score indicates a better diversity and that the generated samples receive patterns from more training shapes. To evaluate the fidelity, we enumerate all parts within one training shape and match each part to the corresponding generated one with the minimum distance. Then we compute the mean distance of these parts as the matching score of one shape. Finally we average the matching scores of all shapes as the MMD score. A lower MMD score indicates that the generation quality of the model is better. Note that both MMD and COV can be computed using either the CD (Chamfer Distance) or EMD (Earth Mover Distance).



To make a fair comparison with G2L, we upsample its generated $(32 \times 32 \times 32)$ voxel maps into $k$ $(32 \times 32 \times 32)$ voxel maps and $6D$ bounding boxes to make it have the same data representation as SAGNet. Then we compute point clouds based on the processed voxel maps and bounding boxes. In testing stage, we generate 1000 shapes to compute for each object class of data. As carefully evaluated in Table~\ref{tab:total_score}, SAGNet achieves the best results in most cases.}

\begin{figure*}[t!]
	\centering
	\includegraphics[width=0.95\linewidth]{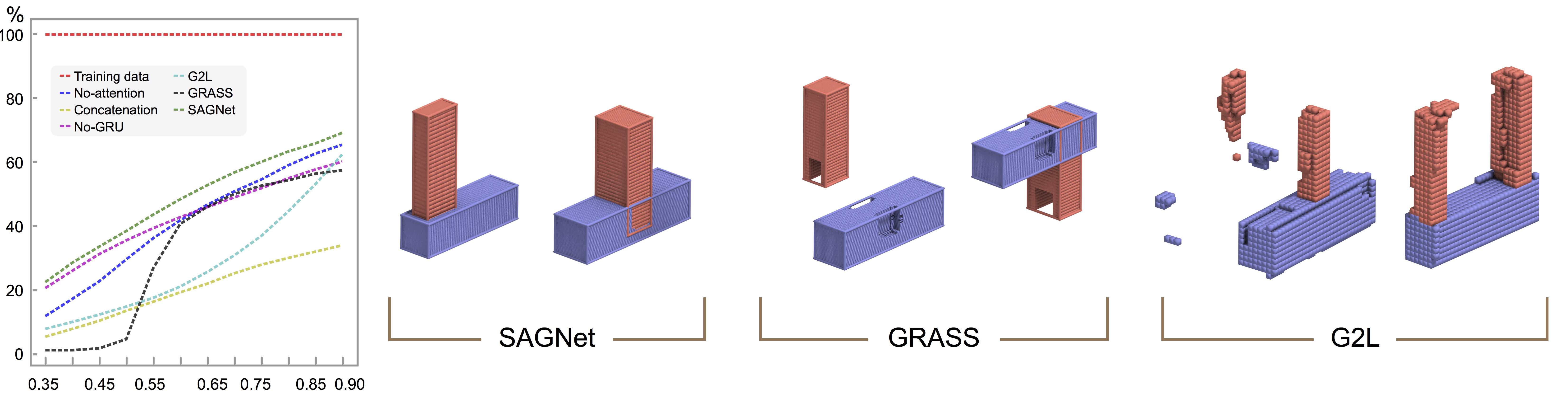}
	\caption{\rs{Cavity analysis and comparison on the synthetic tenon-mortise joints. We present two generated joint shapes by SAGNet, G2L and GRASS models, respectively, in the right. In the left, along the horizontal axis, we provide different thresholds for the fitting accuracy $R$. Along the vertical axis, we provide the percentage of shapes, which have better fitting accuracy than the given thresholds.}}
	\label{fig:toy}
\end{figure*}

\begin{table*}[t!]
	\caption{\rs{Comparisons between SAGNet and baselines and existing models on quantitative scores over synthetic tenon-mortise joints. The results include overlapping score and 3D Inception Score. For both $R_{over}$ and \textit{Inception Score}, the bigger value is better. SAGNet achieves the best results. In contrast, since GRASS processes geometry and structure separately, its overlapping and Inception scores are much lower. Similarly, G2L does not model the relationship between geometry and structure explicitly and so achieves worse results. Please note that G2L has the lowest $R_{o}$ due to its voxel-based representation.}}
	\small
	\centering
	\begin{tabular}{c||ccccccc}
		\hline
		\multirow{2}*{} & \multicolumn{7}{c}{Scores on Synthetic Data} \\\cline{2-8}
		&SAGNet & No-attention & No-GRU & Concatenation & G2L & GRASS  & Training data  \\\hline\hline
		$R_{o}$ & 0.291 & 0.343 & 0.301 & 0.307 & 0.086 & 0.554  & 0.0  \\\hline
		$R_{e}$ & 0.585 & 0.593 & 0.544 & 0.321 & 0.298 & 0.683 & 1.0   \\\hline
		$R_{over}=R_{e}-R_{o}$ & \textbf{0.294} & 0.250 & 0.243   & 0.013 & 0.211 & 0.129  & 1.0   \\\hline		
		Inception Score & \textbf{6.26} & 6.01 & 5.95  & 5.32 & 5.44 & 1.95  &7.98   \\\hline
	\end{tabular}
	\label{tab:toy_example_scores}
\end{table*}

\subsection{Learning the Geometry-Structure Relationships}
\label{ablation}

\paragraph{Symmetry Analysis}
We conduct a part symmetry analysis to evaluate the quality of generated results produced by SAGNet, or baselines, or existing models. Given a generated shape, there are pairs of parts, e.g., the legs of chairs and the wings of airplanes, that are supposed to be symmetric; see Fig.~\ref{Fig:symmetry_score}. We compute the symmetry score for these parts. Specifically, here we focus on the legs of chairs. We perform a mirror reflection of one leg to another. To measure how similar the legs are after the reflection, we use the distance defined in Section~\ref{subsec:generation} as the score. The model that yields lower scores performs better. In Fig.~\ref{Fig:symmetry_score}, we randomly select 1000 training shapes, computing their scores. We provide the percentage of shapes having lower scores than the given thresholds. We find that the training shapes generally have lower scores, showing strong symmetric property.

For each model, we randomly collect 1000 generated chairs and compute their symmetry scores. In Fig.~\ref{Fig:symmetry_score}, we report the scores of SAGNet and other models. SAGNet results in lower scores than those of baselines and existing models, which demonstrates that it can better learn the symmetric property of shapes.

\paragraph{Coplanarity Analysis}
Similarly to symmetry, many objects exhibit coplanarity of parts; see Fig.~\ref{Fig:coplanarity_score}. Here we use the wings (i.e., two fore-wings and two back-wings) of airplanes, measuring their coplanarity to show the quality of the generated shapes. We compute a centroid for each wing of a generated airplane with the corresponding bounding box. Then we use the centroids of two fore-wings and the left back-wing to determine a plane in 3D space. We then compute the Euclidean distance from the centroid of the right back-wing to that plane. Smaller distances imply a better coplanarity property. We select 1000 airplanes from the training data, and compute the centroid-to-plane distance for each airplane. In Fig.~\ref{Fig:coplanarity_score}, we report the percentage of training shapes having smaller distances than given thresholds. The training shapes shows high coplanarity.

We also report the generative models' performances in terms of coplanarity. We randomly generate 1000 airplanes using each model, and compute the centroid-to-plane distance for each airplane. In Fig.~\ref{Fig:coplanarity_score}, we show that SAGNet produces shapes having better coplanarity property than shapes produced by most other models.
Note that GRASS only employs structure information, which is easy to learn, and thus slightly outperforms SAGNet. But our approach accounts for the complex learning of dependencies between structure and geometry, yielding more significant improvement on more challenging generation cases.

\paragraph{Cavity Analysis}
\rs{In order to explicitly show the ability to process and learn the relationship between geometry and structure, we synthesize a dataset as in Fig.~\ref{Fig:toy_example} to train our neural network. Each \emph{tenon-mortise joint} consists of two parts. One part, in blue, has a cavity into which the second part, in red, exactly fits.
There are eight different connection modes according to how one part interacts with the other.
The geometry of the blue, non-convex part is hard to infer just from its bounding box. Only by learning the geometry-structure relationships between the two parts, it is possible to infer the geometry of the non-convex part. The same must hold in order for the generated convex part to exactly fit the cavity.}

We use 10000 such synthetic example shapes to train our framework. Next, we randomly generate 1000 test samples with the trained network and measure how well the convex parts fit into the cavity of the non-convex ones. To quantitatively measure the fitting accuracy, we calculate, $R_{o}$($R_{e}$), the portion of occupancy(empty) voxels of the non-convex part that are overlapped with occupancy voxels of the convex part. With smaller $R_{o}$, there are less occupancy voxels wrongly placed in the convex part. A larger $R_{e}$ means that the convex part fits the cavity of the non-convex part better. Then the score $R = 1 - (R_{e}-R_{o})$ is used to measure how well a generated sample satisfies the dependency. The smaller $R$ indicates better fitting status between the two parts. \rs{Similar to compute MMD and COV, we upsample the generated $(32 \times 32 \times 32)$ voxel maps of G2L into $k$ $(32 \times 32 \times 32)$ voxel maps and $6D$ bounding boxes to make it have the same data representation as SAGNet. We then compare G2L with other models over $R$. Fig.~\ref{fig:toy} shows the performance of each model.} 

\rs{Besides the \emph{Overlapping Score}, we introduce \emph{Inception Score}~\cite{salimans2016improved} to further evaluate. Since each \emph{tenon-mortise joint} can be labeled based on their own connection modes, we are able to train a canonical classifier for them to calculate the \textit{3D Inception Score}. The \textit{Inception Score} was first proposed by Salimans et al.~\shortcite{salimans2016improved}, which is developed to measure the variety and quality of the generated shape set here. In this paper, SO-Net~\cite{li2018sonet} is applied as the classifier. Note that to train this classifier, we should first turn all the training examples with corresponding voxel maps and bounding boxes into point clouds. Besides the 3D coordinates, each point has a label to indicate it belongs to the convex part or non-convex part. At inference stage, we first input all the generated point clouds with their corresponding point labels. Then we compute the confidence of this classifier and variance of the generated classes in a similar manner to G2L~\cite{G2L18}. Note that the shape labels indicate which connection mode one shape belongs to while the point labels refer to that one point is attached to the convex part or non-convex part.

In Table~\ref{tab:toy_example_scores}, we list the exact \emph{3D Inception Score} and \emph{Overlapping Score} produced by SAGNet and baselines and existing models. SAGNet achieves the best performance and significantly outperforms other models thanks to its power of jointly considering the relations between geometry and structure.}


\begin{figure}[t!]
	\centering
	\includegraphics[width=\linewidth]{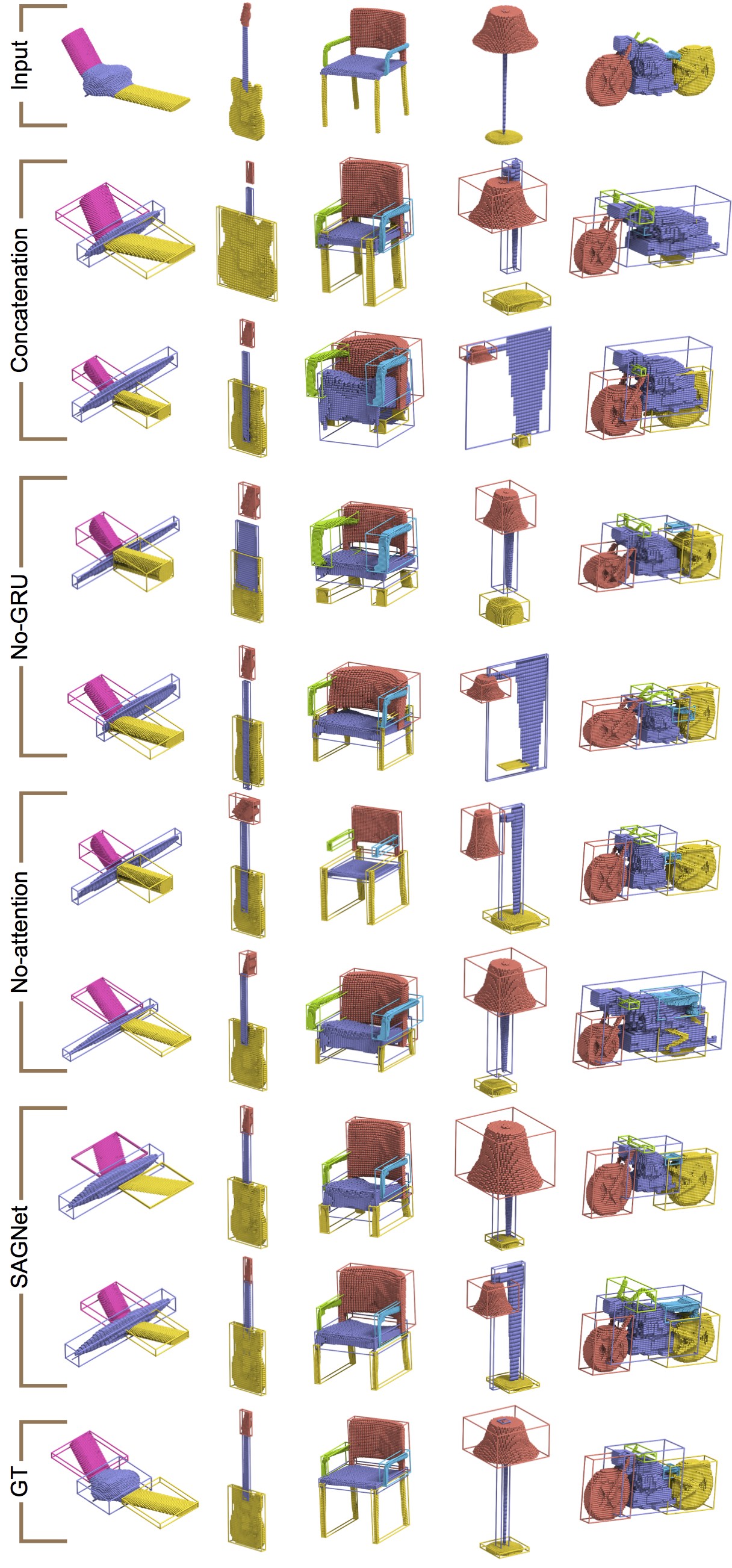}
	\caption{Visual comparison of geometry-to-structure mapping results.}
	\label{Fig:g2s}
\end{figure}

\begin{figure}[t!]
	\centering
	\includegraphics[width=\linewidth]{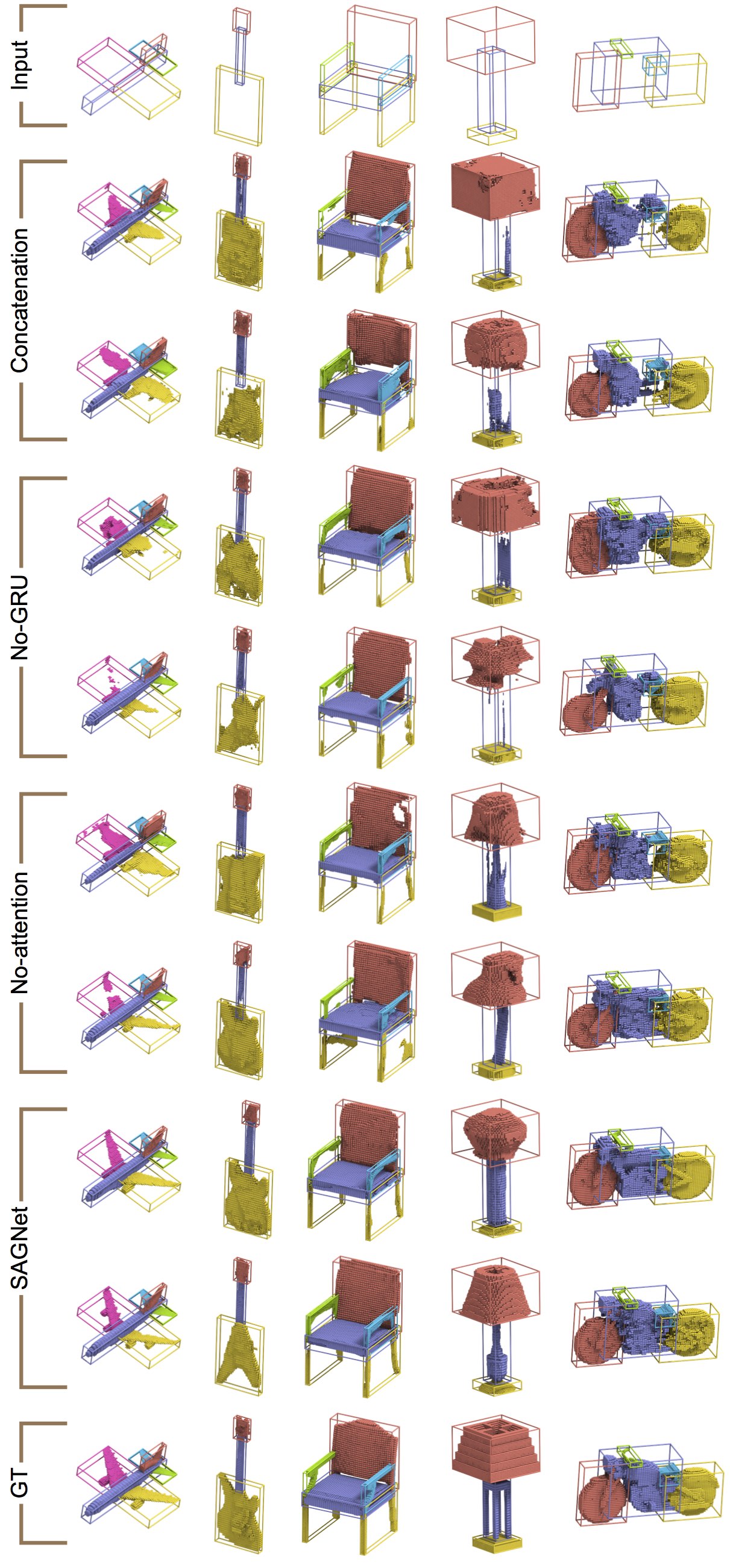}
	\caption{Visual comparison of structure-to-geometry mapping results.}
	\label{Fig:s2g}
\end{figure}

\begin{figure}[t!]
	\centering
	\includegraphics[width=\linewidth]{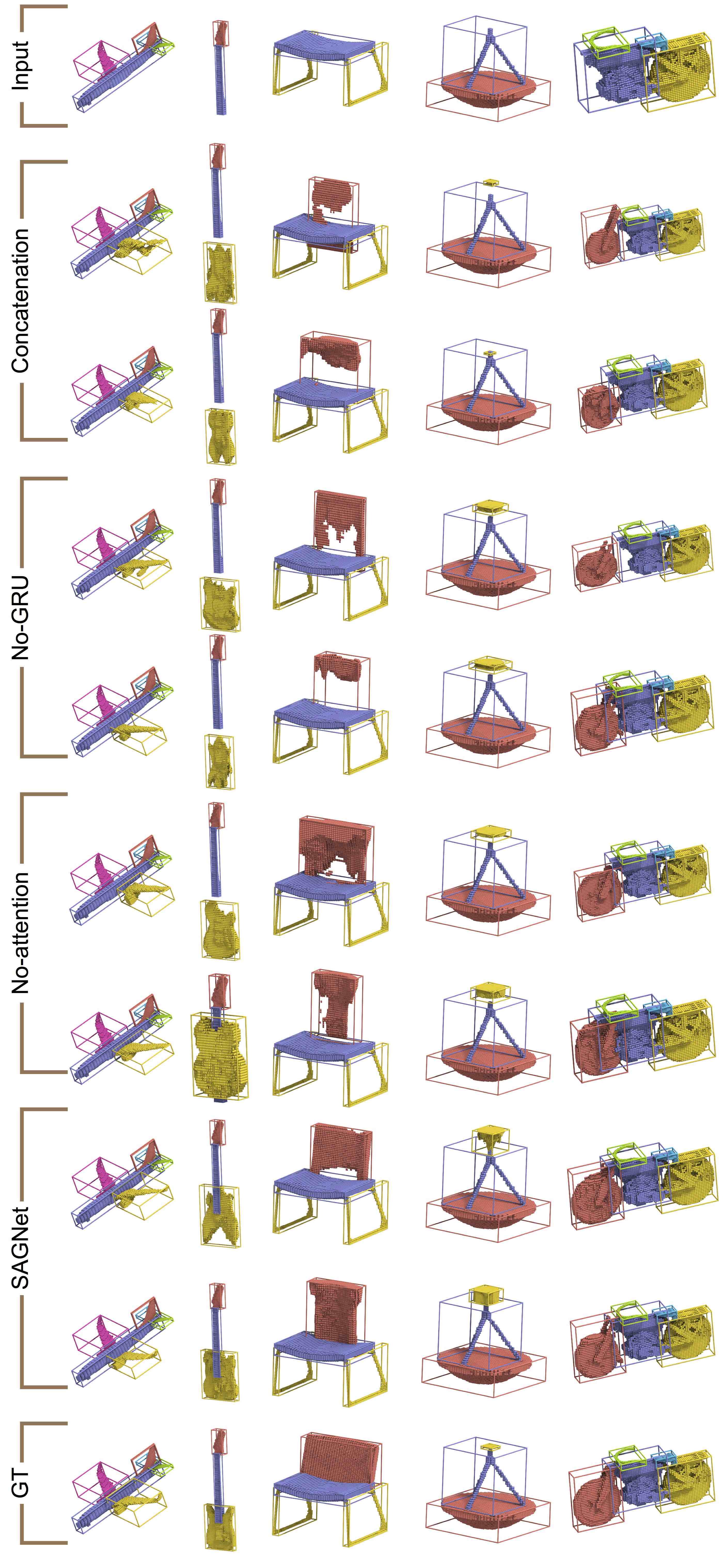}
	\caption{Visual comparison of shape completion results.}
	\label{Fig:completion_result}
\end{figure}

\begin{figure}[t!]
	\centering
	\includegraphics[width=\linewidth]{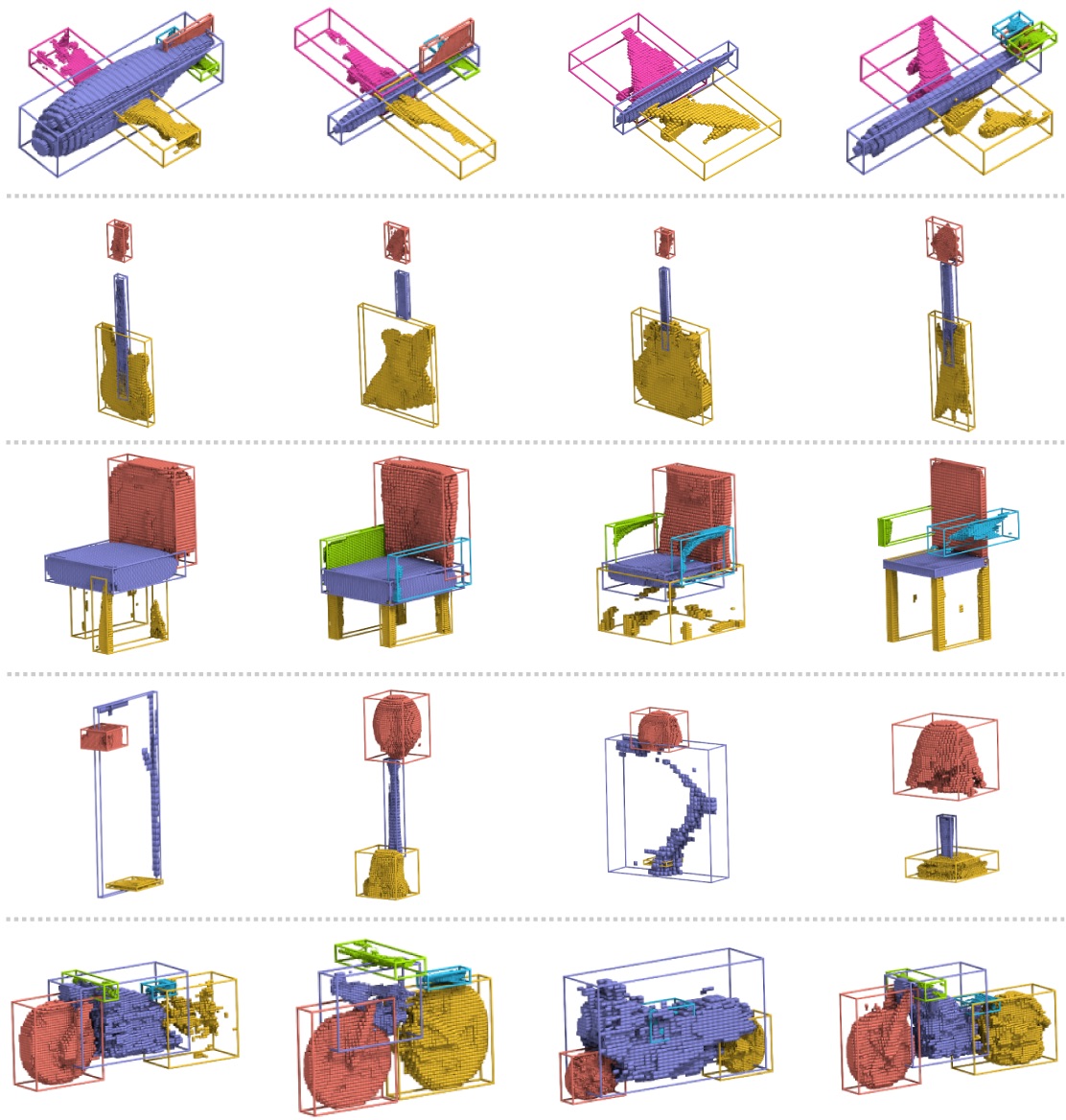}
	\caption{
		Failure examples of SAGNet. Due to the noise and large variation within our training dataset and the learning limitation in current geometry-structure representation, the generated shapes may suffer from disjoint and incompatible parts, geometric detail loss and structure damage.
	}
	\label{Fig:failure_cases}
\end{figure}

\subsection{Geometry-Structure Mapping}
Jointly analyzing the geometry and structural information learns their interdependency. Given geometry information only, a generative model that captures the dependency better is supposed to provide a reasonable inference of the structural information, and vice-versa. Thus we conduct a bidirectional mapping between the geometry and structural information to examine whether SAGNet learns the dependency well; see, e.g., Figs.~\ref{Fig:g2s} and~\ref{Fig:s2g}.

We take all the training data from each class to test the models. Given the voxel maps of all parts only, we randomly initialize the corresponding bounding boxes. Then the voxel maps and bounding boxes are input to the two-branch autoencoder, generating a new set of voxel maps and bounding boxes, which are fed back into the autoencoder. \rs{This is repeated for 300 iterations in order to converge to a marginal distribution~\cite{nash2017shape, rezende2014stochastic}. We compute the Euclidean distance between the ground-truth and generated bounding boxes for each object. The distances for different objects are averaged and listed in Table~\ref{tab:total_score}; check the entry of G2S. We follow a similar procedure to map bounding boxes to voxel maps. The voxel maps are randomly initialized and input to the two-branch autoencoder for 300 iterations. We compute the Chamfer distance using the ground-truth and generated voxel maps, averaging the distances for different object; check the entry of S2G. Clearly, SAGNet outperforms all the three baselines, indicating that it can effectively exchange information to better capture the dependencies between geometry and structure.}



\subsection{Shape Completion}

We further conduct a shape completion task to evaluate the models. Given an object with missing parts, the models should complete these parts by inferring their voxel maps and corresponding bounding boxes. It requires the models to learn the underlying relationship between parts effectively. Given an object, the missing parts are randomly initialized with voxel maps and bounding boxes that are input to the two-branch autoencoder. We follow a similar completion process in~\cite{rezende2014stochastic, nash2017shape}.
The voxel maps and bounding boxes of the missing parts are output by the autoencoder and fed back into it again for 300 iterations to produce the final completion result.

For each class, we take the training data as ground-truth. For each object, we remove some of their parts. The objects with missing parts are then input into SAGNet and baseline models, respectively. To measure the quality of completion, we compute the distance (Section~\ref{subsec:generation}) between the ground-truth parts and the corresponding generated parts. The distances for different objects are averaged. Again, lower score is better. We report the scores of SAGNet and three baseline models in Table~\ref{tab:total_score}. In general, SAGNet achieves lower scores than the baseline models. This is because SAGNet is better equipped to exploit the relationship between parts. In Fig.~\ref{Fig:completion_result}, we compare our shape completion results with the baseline models. SAGNet completes the missing parts with better visual details.

\subsection{Failure Examples}

In Fig.~\ref{Fig:failure_cases}, we provide some examples of failures generated by SAGNet.
Note that high-dimensional voxel maps contribute very complex patterns, increasing the difficulty for learning the representations for 3D shapes.
The learning process inevitably loses geometric and structural details, and the VAE in the proposed model has limited ability to generate fine-grained features.
Similarly, without explicit constraints on the structure definition, the parts of some generated shapes might be disjoint and incompatible to each other on different scales. There is currently no guarantee that all generated shapes are of high quality.

\section{Conclusions and Future Work}

We have presented a network that allows generating 3D shapes with separate control over their geometry and structure. We use weak supervision, in the form of a semantically segmented training set, in order to learn the implicit dependencies between the geometry of parts and their spatial arrangement.
More specifically, we have explicitly demonstrated that the geometry generated in one bounding box, representing one part, is aware of the geometry generated in another bounding box. Since the learned pairwise relationships among the different parts reflect the structure of the shape, we refer to our generative model as structure-aware.

It should be noted that our two-branch autoencoder has similarities with conditional autoencoders in the sense that it encodes information coming from two sources. However, here the two branches learn to extract and intertwine geometry and structure features. This opens up more possibilities for future research. One is to learn other properties in parallel using two separate branches, and intertwine them by a two-way autoencoder. For example, one branch could learn the style of an object and encode it in a feature, while the other branch learns the geometry, and fuse these two features together. Another direction is the development of a $k$-way autoencoder (with $k > 2$), where $k$ properties are learned in parallel using $k$ interconnected branches. The challenge is then to create or collect proper datasets to weakly supervise the learning.

The current training data assumes the objects are segmented into semantic parts. The generative model that we presented does not fully exploit the potential of such data. One can learn more about the geometry of the parts themselves, possibly by employing part-level generators that could potentially generate finer details.

%

In the future, we would like to further explore the fusion of geometry and structure information for 3D shape generation. Currently, we use a fixed numbers of parts that a 3D shape might have, losing the flexibility for modeling various 3D shapes. On the other hand, allowing too many parts may increase the complexity of the learning process. There is still considerable room for improvement of the 3D shape generation quality. In another line, we will explore full representation learning to capture more complex geometry and structure details. It may help in extending the idea of joint analysis of geometry and structure to other types of data, e.g., point clouds and meshes.

\section*{Acknowledgments}
We thank the reviewers for their valuable comments.
This work was supported in parts by National 973 Program (2015CB352501), NSFC (61761146002, 61861130365, 61702338), Guangdong Science and Technology Program (2015A030312015), Shenzhen Innovation Program (KQJSCX20170727101233642), LHTD (20170003), ISF (2366/16), ISF-NSFC Joint Research Program (2472/17), and the National Engineering Laboratory for Big Data System Computing Technology.

\bibliographystyle{ACM-Reference-Format}
\bibliography{SAGNet}

\end{document}